\documentstyle[12pt]{article}
\input epsf

\begin{document}
\baselineskip=15pt
\newcommand{\xx}{{\bf x}}
\newcommand{\yy}{{\bf y}}
\newcommand{\del}{{\partial}}
\newcommand{\bc}{\begin{center}}
\newcommand{\ec}{\end{center}}
\newcommand{\be}{\begin{equation}}
\newcommand{\ee}{\end{equation}}
\newcommand{\bq}{\begin{eqnarray}}
\newcommand{\eq}{\end{eqnarray}}
\newcommand{\x}{{\bf x}}
\newcommand{\p}{\varphi}
\newcommand{\Sc}{Schr\"odinger\,}
\begin{titlepage}
\rightline{DTP 96/65}
\vskip1in
\begin{center}
{\large  Short Distance Properties from Large Distance Behaviour} 
\end{center}
\vskip1in
\begin{center}
{\large
Paul Mansfield and Marcos Sampaio

Department of Mathematical Sciences

University of Durham

South Road

Durham, DH1 3LE, England} 

\begin{center}
{\large 
and Jiannis Pachos\\
Center for Theoretical Physics\\
Massachusetts Institute of Technology\\
Cambridge, MA 02139-4307, USA.}
\end{center}

{\it P.R.W.Mansfield@durham.ac.uk, M.D.R.Sampaio@durham.ac.uk,}
{\it Pachos@ctpa02.mit.edu}
\end{center}
\vskip1in
\begin{abstract}
\noindent
For slowly varying fields the vacuum functional of a quantum field theory
may be expanded in terms of local functionals. This expansion
satisfies its own form of the \Sc equation from which the expansion
coefficents can be found. For scalar field theory in 1+1 dimensions 
we show that this approach
correctly reproduces the short-distance properties as contained in
the counter-terms. We also describe an approximate simplification
that occurs for the Sine-Gordon and Sinh-Gordon vacuum functionals. 
\end{abstract}

\end{titlepage}

\section{\bf Introduction}

Whilst asymptotic freedom has lead to an accurate determination of the
Lagrangian of the Standard Model of particle physics from high energy 
experiments there are few analytical 
tools enabling us to calculate with that Lagrangian at low energies 
where the semi-classical approximation is no longer valid. For example,
the eigenvalue problem for the Hamiltonian of Yang-Mills theory
cannot be solved in a semi-classical expansion because the renormalisation 
group implies that the energy
eigenvalues depend non-perturbatively on the coupling.
Consequently the computation of the hadron spectrum can only be
done numerically. In ordinary quantum mechanics there
are many ways to tackle this problem which are not
widely used in a field theory context, but which,
if suitably generalised might allow non-perturbative 
methods to be developed for field theory.
The oldest of
these is the \Sc representation, (see \cite{Hat}-\cite{else} for
applications to field theory, and \cite{Horiguchi}-\cite{Maeda} for
applications to the Wheeler-de Witt equation). 

\medskip
   In the \Sc representation the vacuum, $|E_0\rangle$, of a scalar 
quantum field theory is represented by the functional $\langle \varphi |E_0\rangle
=\exp W [\varphi]$, where $\langle \varphi |$ is an eigenbra of
the field operator $\hat\phi({\bf x})$ at fixed time, belonging to 
eigenvalue 
$\p(\x )$. In general $W$ is
non-local, but if $\p (\x )$ varies slowly on the scale of the inverse of 
the mass of the lightest
particle,
$m_0^{-1}$,  it can be expanded in terms of local functionals, \cite{Paul1},
for example 
in 1+1 dimensions
\be
W=\int \,dx\, \sum B_{j_0..j_n}\p(x)^{j_0}\p'(x)^{j_1}..
\p^{(n)}(x)^{j_n}.\label{eq:expan}
\ee
The coefficents
$B_{j_0..j_n}$ are constant, assuming translation invariance, and finite as
the ultra-violet cut-off is removed, \cite{Sym}. 
Particle structure is characterised by length scales smaller 
than $m_0^{-1}$, so this simplification in $W[\p]$
does not appear useful, however, a knowlege of this local expansion is
sufficent to reconstruct $W$ for arbitrary $\p$,\cite{Paul2}. 
This is because if
$W[\p]$ is evaluated for a scaled field,
$\p_s(x)\equiv\p (x/\sqrt s)$, it extends to an analytic function of $s$
with cuts restricted to the negative real axis, so that Cauchy's theorem 
can be used to relate the large-$s$ behaviour (when $\p_s$ is slowly
varying) to the $s=1$ value:
\be
W[\p]=\lim_{\lambda\rightarrow\infty}{1\over 2\pi i}\int_{|s|=\infty}
{ds\over s-1}e^{\lambda(s-1)}W[\p_s].
\label{eq:resss}
\ee
The exponential term removes the contribution of the cut as $\lambda
\rightarrow\infty$.
The vacuum functional satisfies the \Sc equation from which the coefficents
$B_{j_0..j_n}$ can be found in principle, however care must be exercised
because this equation depends explicitly on short-distance effects
via the cut-off, whereas the local expansion is only valid for
fields characterised by large length-scales, so we cannot simply substitute
(\ref{eq:expan}) into the \Sc equation and expect to be able to satisfactorily take
the limit in which the cut-off is removed. However, we can again exploit 
Cauchy's theorem to construct a version of the \Sc equation that acts directly
on the local expansion by considering the effect of a scale-transformation
on the cut-off, as well as on the field,\cite{Paul1}. The Hamiltonian with 
a scaled cut-off acting on the vacuum functional evaluated for the scaled field
again extends to an analytic function with cuts on the negative real axis.
This enables the limit in which the short-distance
cut-off is taken to zero to be expressed in terms of large-distance 
behaviour described by the local expansion for 
$W[\p]$. This leads to an infinite set of algebraic equations for the
coefficents $B_{j_0..j_n}$. By truncating the expansion 
the \Sc equation offers the possibility of solution beyond
perturbation theory in the couplings, however, before this is
attempted it is essential to show that this formulation
is capable of reproducing the results that can be obtained using 
the standard approach of the semi-classical expansion and 
Feynman diagram perturbation theory. In particular, since the
method consists of building states out of their large distance properties
it is important to show that it gets right the short-distance behaviour
as contained in the counter-terms of the Hamiltonian.
The purpose of this paper is to demonstrate that this short-distance
behaviour is correctly reproduced by our approach to the \Sc equation 
in which we build the vacuum state from its large-distance behaviour,
and that, at least to low orders, the resulting local expansion coincides
with the Feynman diagram calculation of the vacuum functional.

\section{\bf Semi-classical analysis of local expansion}

The classical Hamiltonian of $\p^4$ theory is
$\int dx ({1\over 2}\left( \pi^2+\varphi^{\prime
2}+m^2\varphi^2\right) +{g\over 4!}\varphi^4)$ in 1+1 dimensions. 
In the \Sc representation
the canonical momentum is represented by functional differentiation
$\hat \pi=-i\hbar\delta/\delta\p(x)$, so the kinetic term leads to
 the product of two
functional derivatives at the same point which we regulate by introducing 
a momentum cut-off $p^2<1/\epsilon$. The couplings must consequently be 
renormalised so that the Hamiltonian has a finite action on the vacuum.
Writing the vacuum functional as $\exp\,(W[\p]/\hbar)$ gives the 
\Sc equation as $\lim_{\epsilon\downarrow 0} F_\epsilon[\p]=0$ where
\be
F_\epsilon[\p]=
-{\hbar\over 2}\Delta_\epsilon W+
\int dx\left({1\over 2}\left(-\left({\delta W\over\delta\p}\right)^2
+\p'^2+ M^2(\epsilon)\p^2\right)
+{g\over 4!}\p^4 -{\cal E}(\epsilon)\right)
\ee
and
\be
\Delta_\epsilon=\int dx\,dy\int_{p^2<1/\epsilon}{dp\over 2\pi}e^{ip(x-y)}{\delta^2
\over\delta\p(x)\delta\p(y)}
=\int_{p^2<1/\epsilon}{dp\,2\pi}{\delta^2
\over\delta\tilde\p(-p)\delta\tilde\p(p)},
\ee
where $\tilde\p(p)=\int dx \p(x)\exp(-ipx)$.
In perturbation theory the only divergent diagrams
with external legs are tadpoles, and these can be removed by
normal ordering the Hamiltonian. This enables the $\epsilon$-dependence
of the parameters to be calculated exactly as

\be
M^2(\epsilon)=M^2+\hbar\delta M^2-\hbar{g\over 4}\int_{p^2<1/\epsilon} {dp\over 2\pi}
{1\over\sqrt{p^2+M^2}},
\label{eq:M}
\ee

\be
{\cal E}(\epsilon)=\delta{\cal E}+{\hbar\over 2}\int_{p^2<1/\epsilon}
{dp\over 2\pi}\left(\sqrt{p^2+M^2}+{M^2(\epsilon)-M^2\over
2\sqrt{p^2+M^2}}\right)+{g\hbar^2\over 32}\left(\int{dp\over 2\pi}{1\over
\sqrt{p^2+M^2}}\right)^2
\label{eq:ensub}
\ee
where $M^2,\delta M^2,\delta{\cal E}$ and ${\cal E}$ remain finite 
when the cut-off is removed.
The ambiguity in the choice of counterterms represented by
$\delta M^2$ and $\delta{\cal E}$ is resolved, as usual, by renormalisation conditions.
We shall soon see that there is a natural way to do this in the present context.
If $F_\epsilon [\p]$ is evaluated for a $\p$ whose Fourier
transform is non-zero only for momenta less than $m_0$ it will reduce to
a sum of local functionals of $\p$, 
\be
F_\epsilon [\p]=
\int \,dx\, \sum f_{j_0..j_n}(\epsilon)\p(x)^{j_0}\p'(x)^{j_1}..
\p^{(n)}(x)^{j_n}\label{eq:Fexpan}
\ee
The expansion functions, 
$\p^{j_0}\p'^{j_1}\p''^{j_2}..$, are 
related by partial integration so we can specify a linearly independent basis
by insisting that the power of the highest derivative be at least two,
and we will assume parity and $\p\rightarrow-\p$ invariance, restricting
both the total number of $\p$ and the total number of derivatives in the
expansion functions appearing in (\ref{eq:expan}) and (\ref{eq:Fexpan}) to be even.  
It is important to note that (\ref{eq:Fexpan}) is not the 
same expression that would be obtained by acting with $\Delta_\epsilon$
on the local expansion (\ref{eq:expan}), because the former correctly
includes differentiation with respect to the Fourier modes 
of $\p$ with momenta in the range $m_0^2<p^2<1/\epsilon$ absent from the
second.
Now if we scale the cut-off $(\Delta_{s\epsilon}W)[\p_s]$ extends to an 
analytic function in the complex $s$-plane with singularities only
on the negative real axis, \cite{Paul1},the same is true of $M^2(s\epsilon)$
and ${\cal E}(s\epsilon)$, and consequently of the coefficents
of the linearly independent expansion functions,
$f_{j_0..j_n}$ in (\ref{eq:Fexpan}), so the contour integral
\be
I_{j_0..j_n}(\lambda)={1\over 2\pi i}\int_{|s|=\infty}{ds\over s}e^{\lambda s}
\sqrt{\pi\lambda s}f_{j_0..j_n}(s\epsilon)\label{eq:I}
\ee
can be calculated by collapsing the contour
to a small circle about the origin and a contour
along the cut on the negative real axis.
The contribution from the circle about the origin
is controlled by the small $\epsilon$ behaviour of
$f_{j_0..j_n}(\epsilon)$. As $\epsilon\rightarrow 0$
this vanishes due to the \Sc equation, 
and in perturbation theory the Feynman diagram expansion
gives an asymptotic expansion of $f_{j_0..j_n}(\epsilon)$
in positive powers of $\sqrt\epsilon$. The inclusion of
$\sqrt{\pi\lambda s}$ in (\ref{eq:I}) ensures that
the contribution from the origin will be of order
$1/\lambda$ rather than $1/{\sqrt\lambda}$. 
For large $|s|$ the scaled field $\p_s$
is slowly varying and the scaled cut-off $1/(s\epsilon)$ is less
than $m_0$ so $(\Delta_{s\epsilon}W)[\p_s]$ can now be calculated by
acting with $\Delta_{s\epsilon}$ directly on the local expansion of $W$, 
(\ref{eq:expan}). Furthermore as the real part of $\lambda$
tends to infinity the contribution from the cut tends to zero due to
the $\exp(\lambda s)$ factor, (the contribution from that part of the cut 
for which $|s|$ is large
is again given by the local expansion and seen to be suppressed
as $\lambda\rightarrow\infty$). Thus the \Sc equation leads
to an infinite set of algebraic equations $\lim_{\lambda
\rightarrow\infty}I_{j_0j_1..j_n}(\lambda)=0$ where

\bq
&&I_0=-\bar{\cal E}(\lambda)-
\hbar{\sqrt\lambda\over\sqrt\pi}\left(B_2+{B_{0,2}\lambda\over 3}
+{B_{0,0,2}\lambda^2\over 10}+..\right)\nonumber\\
&&I_2={\bar M^2(\lambda)\over 2}-{2B_{2}^2}
-\hbar{\sqrt\lambda\over\sqrt\pi}\left(6 B_4+{B_{2,2}\lambda\over 3}
+{B_{2,0,2}\lambda^2\over 10}+..\right)\nonumber\\
&&I_4={g\over 4!}-{8B_2B_4}
-\hbar{\sqrt\lambda\over\sqrt\pi}\left(15 B_6+{B_{4,2}\lambda\over 3}
+{B_{4,0,2}\lambda^2\over 4}+..\right)\nonumber\\
&&I_{0,2}={1\over 2}-{4B_2B_{0,2}}
-\hbar{\sqrt\lambda\over\sqrt\pi}\left( B_{2,2}+2B_{0,4}\lambda+
{4B_{2,0,2}\lambda\over 3}
+..\right)\label{eq:form}
\eq
and
\be
\bar{\cal E}(\lambda)={1\over 2\pi i}\int_{|s|=\infty}{ds\over s}e^{\lambda s}
\sqrt{\pi\lambda s}{\cal E}(s)=\sum_0^\infty \hbar^n\bar{\cal E}(\lambda)^{\hbar^n}
\ee

\be
\bar M^2(\lambda)={1\over 2\pi i}\int_{|s|=\infty}{ds\over s}e^{\lambda s}
\sqrt{\pi\lambda s}M^2(s)=M^2+ \hbar {\bar M}^2(\lambda)^{\hbar}
\ee
As the product $s\epsilon$ now plays the r\^ole of cut-off,
rather than $\epsilon$ alone, we have taken
$\epsilon$ to be finite and equal to unity. We will now choose renormalisation 
conditions. Note that the counter-terms only enter $I_0$ and $I_2$.
If these are fixed then the above equations determine the coefficents
$B_{j_0,.,j_n}$ and the energy eigenvalue,
$\cal E$, which are themselves finite as the cut-off is removed.
Alternatively we could instead choose the values of two of these
quantities, $B_2$ and $\cal E$ for example, 
and then think of the equations $I_0=0$ and $I_2=0$ as 
determining the counter-terms. So we will take $B_2=-M/2$,
which is its classical value, and ${\cal E}=0$ as our
renormalisation conditions. The advantage of 
imposing the renormalisation conditions on ${\cal E}$ and $B_2$ is that
we are free to solve (\ref{eq:form}) for the remaining $B_{j_0,..,j_n}$ without
first computing the $\lambda$-dependence of the counter-terms
which in a more general context can only be done in perturbation theory.

    The equations (\ref{eq:form}) may be solved in the usual semi-classical 
approach in which we expand the coefficents as
$B=\sum\hbar^n B^{\hbar^n}$,
by first ignoring the terms proportional to
$\hbar$. Although the resulting equations are quadratic in
the $B_{j_0..j_n}$ they are readily solved by starting
with the coefficents of local functions of the lowest
dimension and number of $\p$, giving at tree-level
\bq
&&W_{\rm tree}=\int dx\Biggl(
-{1\over 2}\p^2-{1\over 4}\p'^2+{1\over 16}\p''^2-{1\over 32}
\p'''^2+{5\over 256}\p''''^2
-{1\over 96} g\p^4 +{1\over 64} g\p^2\p'^2\nonumber\\ &&\,\,\,\,\,\,
-{1\over 128} g\p^2\p''^2
+{1\over 256} g\p'^4+{5\over 1024}
 g\p^2\p'''^2-{3\over 256} g\p\p''^3
-{31\over 1024} g\p'^2\p''^2\nonumber\\
&&\,\,\,\,\,\,-{7\over  2048} g\p^2\p''''^2
+{41\over 1024} g\p\p''\p'''^2
+{75\over 2048} g\p'^2\p'''^2-{93\over 4096} g\p''^4+..\Biggr)
\label{eq:tree}
\eq
where we have chosen our mass-scale so that $M=1$.
Particular tree-level coefficents that will be of use are
\be
B_{0,0,..,0,j_n=2}^{1} = -\frac{1}{2}  
\left( \begin{array}{c}   1/2\\
                           n
       \end{array}
\right) \ \ \ 
\label{Clas.Bn} 
\ee
and
the coefficients $B_4^1 , B_{2,2}^1 , B_{2,0,2}^1 , \ldots , B_{2,0\ldots0,j_n=2}^1$. 
To simplify the following formulae we re-name some of the coefficents. Firstly
let $B^1_2\equiv b_0/2,B_{0,0,..,0,j_n=2}^1 \equiv b_n, n=1,2,..$
and $B_4^1\equiv c_0/6 , B_{2,2}^1 , B_{2,0,2}^1 , \ldots , 
B_{2,0\ldots0,j_n=2}^1\equiv c_n$ then the tree-level contribution to
the equations 
$I_4 = 0 , I_{2,2} = 0 , I_{2,0,2} = 0 , \ldots , I_{2,0\ldots0,2} = 0 $  
can be written as
$$
b_0 c_1 + b_1 c_0 = 0
$$
$$
b_0 c_2 + b_1 c_1 + b_2 c_0 = 0 
$$
$$
b_0 c_3 + b_1 c_2 + b_2 c_1 + b_3 c_0 = 0
$$
\be
etc 
\label{bc}
\ee
which, in turn, may be expressed  as the vanishing of
each coefficent of $z$ in
\be 
\left( \sum_{n=0}^{\infty} b_n z^n \right)\left(
\sum_{m=0}^{\infty} c_m z^m\right) - b_0 c_0 =0,
\label{bc-series}
\ee
We can solve for the
$c_n$ in (\ref{bc-series}) to give 

\be
B_{2,0,..,0,j_n=2}^{1}
=-{g\over16}\left|
\begin{array}{ccccc}
B^{1}_{0,2} & B^{1}_{0,0,2} & .. & B^{1}_{0,0,..,j_{n-1}=2} &     
B^{1}_{0,0,..,j_n=2}      \\
-1 & B^{1}_{0,2} & .. &  B^{1}_{0,0,..,j_{n-2}=2} &  B^{1}_{0,0,..,j_{n-1}=2}\\
0 & -1 & .. &  B^{1}_{0,0,..,j_{n-3}=2} & B^{1}_{0,0,..,j_{n-2}=2} \\
.. & .. & .. & .. & .. \\
0 & 0 & .. & -1 & B^{1}_{0,2}
\end{array}
\right|
\label{eq:Clas.Cn}
\ee
We can find in a similar fashion the coefficients 
$B_6^1\equiv f_0/15,B_{4,0 \ldots 0,j_n=2}^1 \equiv f_n,n=1..$. They are 
determined by the tree level equations $I_6 = 0, I_{4,2} = 0 , I_{4,0,2} = 0, $ etc. 
Redefining $B_2^1 \equiv \frac{1}{3} b_0$,
$B_4^1 =
\frac{1}{8} c_0$  allows us to write those equations as   
\be
\left( \sum_{n=0}^{\infty} c_n z^n \right)^2 + 2 \left(
\sum_{m=0}^{\infty} b_m z^m\right) \left (\sum_{l=0}^{\infty} 
f_l z^l\right)-(c_0)^2 - 2 b_0 f_0=0
\ee
where each coefficent of $z$ must separately vanish. 
If we set $\left( \sum_{m=0}^{\infty} b_m z^m\right)^{-1} =
\sum_{m=0}^{\infty} \beta_m z^m$ as well as 
$\left( \sum_{n=0}^{\infty} c_n z^n \right)^2 = \sum_{n=0}^{\infty}
\gamma_n z^n$ we can formally write, after substituting 
back the original values of $b_0, c_0$ and  $f_0$ ,
\be
f_n = \frac{1}{2} \left( (c_0)^2 + 2 b_0 f_0 \right) \beta_n - \frac{1}{2} 
\sum_{k=0}^{n} \beta_k \gamma_{n-k} = - \frac{1}{2} \left( \frac{1}{27648} \beta_n +  
\sum_{k=0}^{n} \beta_k \gamma_{n-k} \right) \ , 
\ee 
$n \ge 1$ . Using the formulae for inversion and product of power series from the 
mathematical literature \cite{math}, we can
calculate all the $B_{4,0 \ldots 0,j_n=2}^1$
\bigskip

The order-$\hbar$ corrections  are
obtained by substituting the tree-level results into the
previously ignored order-$\hbar$ term in the \Sc equation
and treating this as a perturbation to the classical equation.
We want to use this to show that our large-distance expansion
correctly gives the short-distance behaviour as contained in
the divergent mass and energy subtractions $\bar {\cal E}(\lambda)$
and $\bar M^2(\lambda)$, which occur only in $I_0$
and $I_2$.  We first study $I_2$. Using (\ref{eq:Clas.Cn}) we get
the $O(\hbar)$ expression 

\be
I_2^\hbar (\lambda)={\bar M^2(\lambda)^\hbar\over 2}+2B_{2}^{\hbar}
-g{\sqrt\lambda\over\sqrt\pi}\left({1\over 16}-{\lambda\over 192}
+{\lambda^2\over 1280}-{5\lambda^3\over 43008}+..\right)
\ee
This vanishes when $\lambda\rightarrow\infty$,
but we get a good approximation if we truncate the series
and take $\lambda$ as large as the truncation will allow, i.e.
small enough for the first neglected term to be insignificant.
Since $I(\lambda)$ is of order $1/\lambda$ for large $\lambda$ 
the accuracy of this approximation is greatly improved if we 
perform a further contour integration, amounting to 
a re-summation of the series in $\lambda$.
Observe that substituting $\lambda=1/{\sqrt{ s}}$ in $I 
(\lambda) $ gives a function that is analytic in $ s$ with
a cut on the negative real axis that we wish to evaluate
as $ s$ tends to zero from real positive values,
so we define

\be
\tilde I(\lambda)={1\over 2\pi i}\int_{| s|=\infty}
{d s\over  s}e^{\lambda^2  s}
\sqrt{\pi\lambda^2  s}\,I(s^{-1/2})\label{eq:tildeI}
\ee
for which $\lim_{\lambda\rightarrow \infty}\tilde I(\lambda)
=0$. Thus
\be
\tilde I_0^\hbar (\lambda)={\delta M^2+\tilde M^2(\lambda)^\hbar\over 2} + 2B_{2}^{\hbar}
-gS(\lambda)
\label{eq:io}
\ee
where
\be
S(\lambda)={\sqrt\lambda\over\sqrt\pi}\left({1\over 16\Gamma(3/4)}-{\lambda\sqrt 2
\Gamma(3/4)\over 96\pi}
+{\lambda^2\over 960 \Gamma(3/4)}
-{\lambda^3\sqrt 2
\Gamma(3/4)\over 5376\pi}+
..\right) \label{eq:series}
\ee
The terms in $S(\lambda)$ now decrease more rapidly than the corresponding
terms in $I(\lambda)$.
Since $I(1/\sqrt s )$ behaves asymptotically as $\sqrt s$ for
small $s$
this re-summation has the effect of eliminating the leading term
so that $\tilde I(\lambda)$ is now of order $1/\lambda^2$.
Further re-summations are only efficacious given a 
sufficent number of terms in the truncated series for the
extra gamma-functions in the coefficents to be noticeable.
\begin{figure}[h]
\unitlength1cm  
\begin{picture}(14,10)
\put(10,2.5){$-{\tilde M^2(\lambda)^\hbar/ 2g}$}
\put(9,7.5){$S(\lambda)$} 
\put(3.5,5){ $S(\lambda)-{\tilde M^2(\lambda)^\hbar/ 2g}$ }
\epsfysize=13.5cm
\epsffile{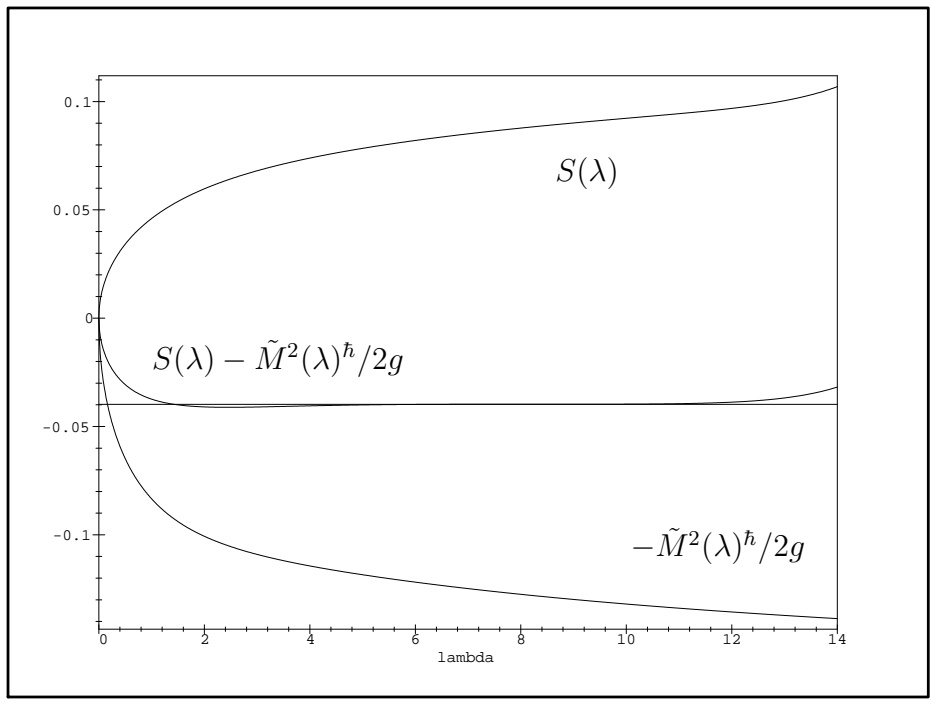} 
\end{picture}
\caption[mass]{\label{mass} 
The mass subtraction}
\end{figure}

\begin{figure}[h]
\unitlength1cm  
\begin{picture}(14,10)
\put(11.7,8.4){$S_{13}$}
\put(8.3,8.4){$S_{9}$}
\put(6.2,8.4){$S_{5}$}
\put(11.5,2){$S_{12}$} 
\put(8,2){$S_8$}
\put(4,2){$S_4$}
\epsfysize=13.5cm
\epsffile{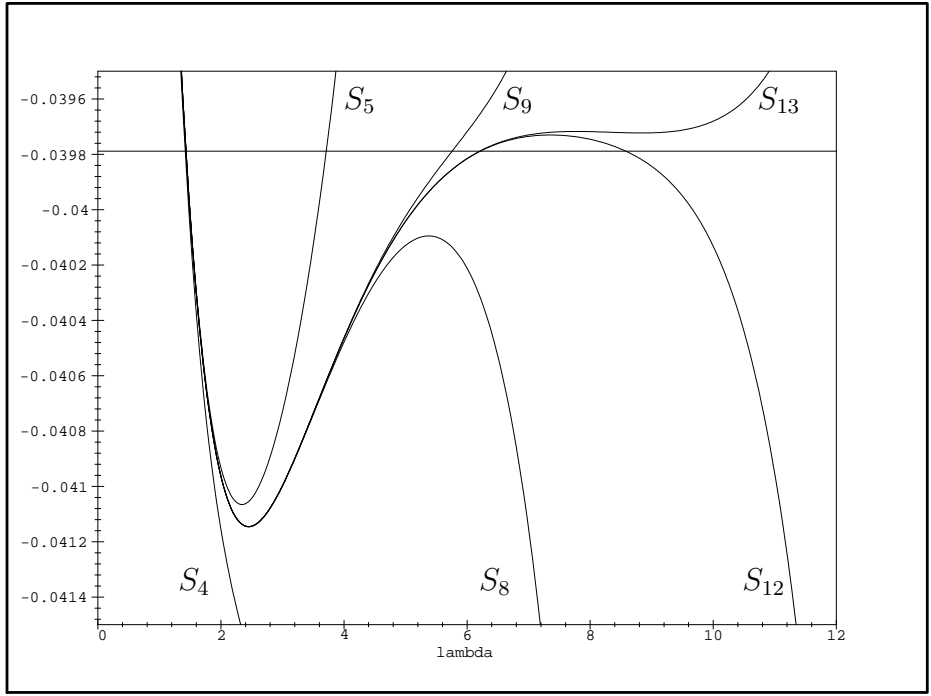} 
\end{picture}
\caption[series]{\label{series} 
Truncating $S(\lambda)$}
\end{figure}

In fig. (\ref{mass})
we plot, the series
$S(\lambda)$ truncated to 13 terms, $-{\tilde M^2(\lambda)^\hbar/ (2g)}$,
their sum, and
the limit of this sum as $\lambda\rightarrow\infty$, (which we obtain exactly
in the next section as $-1/(8\pi )\simeq
-0.0398$).
Clearly neither $S(\lambda)$ nor $-{\tilde M^2(\lambda)^\hbar/ (2g)}$
are constant for large $\lambda$ but their sum is, to a good approximation
for $\lambda>2$. This shows that our large-distance expansion
correctly reproduces the short-distance effects encoded in
$M^2(\epsilon)^\hbar$. The departure from this constant value for 
$\lambda>11$ is due to the error involved
in truncating $S(\lambda)$ to 13 terms. 
If we denote by $S_n$ the series truncated to $n$ terms 
minus ${\tilde M^2(\lambda)^\hbar/ (2g)}$ then
in fig. (\ref{series}) we have shown $S_n$ for $n=4,5,8,9,12,13$.

Each truncation
provides a good approximation to $S(\lambda)$ up to a value 
of $\lambda$ which is large enough 
for the highest order term to be a significant fraction of the whole.
Taking this to be one per cent gives an estimate of $S(\infty)$
with an error that ranges from three per cent (five terms)
to half a per cent (13 terms). 

To check that our large distance expansion correctly reproduces the
energy subtraction we need the $O(\hbar)$ part of $W[\p]$ 
that is quadratic in $\p$. We obtain this from the equations
$I_{0,2}=0,I_{0,0,2}=0,..$, having imposed the renormalisation condition
$B_2=-M/2$. We use the re-summation described earlier,
truncate the series in $\lambda$ so that they include contributions from
coefficents of functionals of $\p$ of dimension less than 26,
and take $\lambda$ so that the last incuded term is one per cent
of the value of the truncated series. We also use Stieltje's
trick of halving the contribution of the last included term to 
improve the accuracy of the approximation \cite{St}.
This gives the estimate 

\bq
&&
W_2^\hbar=
{g\over 1000}\int dx\Bigg({6.64\p'^2}-{6.02\p''^2}
 +{ 5.40\p'''^2 }-{ 4.91\p''''^2}+{ 4.54\p^{(5)2} }\nonumber\\
&&\,\,\,\,\,\,
-{ 4.24\p^{(6)2}}
 +{ 4.01\p^{(7)2} }-{3.79 \p^{(8)2} }+{ 3.58\p^{(9)2} }
 -{3.34 \p^{(10)2}}+..\Bigg)
\eq
\vfill
\eject
In the next section we obtain $W_2^\hbar$ exactly. Rounding
the exact results to three significant figures gives
\bq
&&
W_2^\hbar=
{g\over 1000}\int dx\Bigg({6.63\p'^2}-{5.97\p''^2}
 +{ 5.33\p'''^2 }-{ 4.84\p''''^2}+{ 4.45\p^{(5)2} }
\nonumber\\
&&\,\,\,\,\,\,\,
-{ 4.14\p^{(6)2}}
 +{ 3.89\p^{(7)2} }-{3.68 \p^{(8)2} }+{ 3.50\p^{(9)2} }
 -{3.34 \p^{(10)2}}+..\Bigg)
\eq
which shows that our approximate results are good to a few per cent.

\begin{figure}[h]
\unitlength1cm  
\begin{picture}(14,10)
\put(8.3,8.4){$A$}
\put(8,2){$B$}
\put(4,7){$C$}
\epsfysize=13.5cm
\epsffile{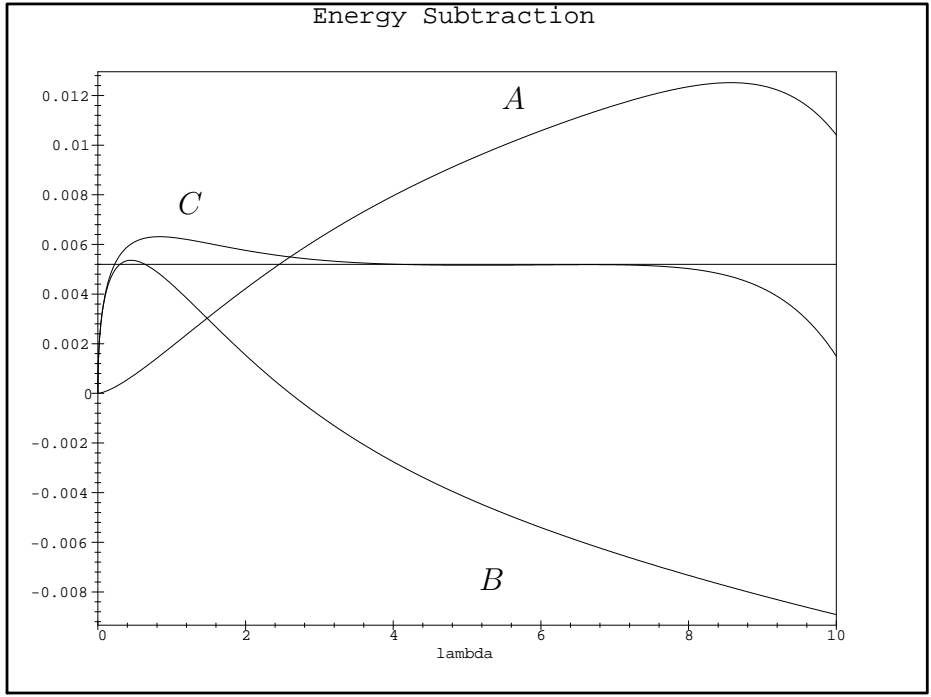} 
\end{picture}
\caption[3]{\label{3} The energy subtraction}
\end{figure}

Figure (\ref{3}) shows the effect of substituting this estimate into
the $O(\hbar^2)$ contribution to
$I_0$. The top curve, $A$, is the estimate of the re-summation of the 
series in $\lambda$, whilst the bottom curve, $B$, is the $O(\hbar^2)$
contribution to the re-summation of $\bar{\cal E}(\lambda)$
evaluated using (\ref{eq:ensub}) with $\delta M^2=g/(4\pi)$.
Neither of these curves appears to tend to a constant for large
$\lambda$ whereas their sum, represented by the middle curve, $C$,
provides a good approximation to a constant value 
for $\lambda$ larger than four until $\lambda$ is sufficently
large that the approximation of the infinite series   
by just ten terms breaks down.
The straight line in the figure is the value $0.0052$ 
which would be obtained by truncating the series at fifty 
terms using the expression for $W^\hbar_2$ we find in the next section.

\medskip
  Having seen that our large distance expansion successfully
reproduces the short-distance effects contained in the 
counter-terms of the Hamiltonian we turn to the one-loop
evaluation of the $B_{j_0,..,j_n}$ coefficents corresponding to
higher numbers of fields. Begin with the coefficents of
local functions containing four fields. Rounding the tree-level
result to three significant figures gives
\bq
&&W^1_4={g\over 1000}\int dx \Biggl(-10.4\p^4+15.6\p^2\p'^2
+3.91\p'^4         \nonumber\\
&&\,\,\,\,\,-7.81\p^2\p''^2   -30.3\p'^2\p''^2 -11.7\p\p''^3+4.88\p^2\p'''^2
-22.7\p''^4\nonumber\\
&&\,\,\,\,\,\quad\quad\quad\quad\quad\quad\quad\quad+36.6\p'^2\p'''^2+
40.0\p\p''\p'''^2-3.42\p^2\p''''^2+..
\Biggr)
\eq
Estimating the $O(\hbar)$ contribution in the same way that we estimated
$W_2^\hbar$ gives
\bq
&&W^\hbar_4={g^2\over 10000}\int dx \Biggl(4.02\p^4-20.0\p^2\p'^2
-7.96\p'^4         \nonumber\\
&&\,\,\,\,\,17.4\p^2\p''^2   +83.8\p'^2\p''^2 +37.6\p\p''^3-15.6\p^2\p'''^2
+87.7\p''^4\nonumber\\
&&\,\,\,\,\,\quad\quad\quad\quad\quad\quad\quad\quad-129\p'^2\p'''^2+
-164\p\p''\p'''^2+14.0\p^2\p''''^2+..
\Biggr)
\eq
There are two things to note. Firstly there is a proliferation of local
functionals of the same dimension and number of $\p$ as these increase.
So, for example, there is a unique local functional with just two $\p$ 
for any dimension,
but there are two hundred and seven with twelve $\p$ and dimension twelve.
Secondly the ratio of the $O(\hbar)$ corrections to any two 
coefficents of functionals containing the same number of $\p$ and the same dimension is approximately the same as the ratio of the tree-level
values. For example the ratio of the $O(\hbar)$ coefficents of $\p^2\p''^2$
and $\p'^4$ is $-17.4/7.96\simeq-2.19..$ whereas the ratio of the
corresponding tree-level values is exactly $-2$. Given that our estimate
is probably only good to a few per cent it is not clear 
at this stage whether the one-loop ratios are exactly equal to 
the tree-level ratios, but we will investigate this with greater 
accuracy in the next section. We will now compare these results  with those obtained 
by solving the \Sc equation without first expanding
in terms of local functions.

\section{\bf Direct Semi-Classical Solution}

It is straightforward to solve the \Sc equation $\lim_{\epsilon\downarrow
0}F_\epsilon[\p]=0$ without
resorting to the local expansion, at least  for low orders of an expansion in $\p$ and
$\hbar$,\cite{Hat}. This turns out to be remarkably efficent compared to the
Feynman diagram expansion which we describe in the next section.
Expand $W[\p]$ as
\be
W[\p]=\sum_{n=1}^\infty \int dp_1..dp_{2n}\tilde\p(p_1)..\tilde\p(p_{2n})
\Gamma_{2n}(p_1,..,p_{2n})\delta(p_1+..+p_{2n})
\ee
where the $\Gamma$ are unknown functions.
Then we can write $\Delta_\epsilon W[\p]
=\sum \Delta\Gamma_{2n}$ where $\Delta\Gamma_{2n}$ is
\be
\int_{q^2<1/\epsilon}2\pi
dq \int dp_3..dp_{2n}2n(2n-1)\tilde\p(p_3)..\tilde\p(p_{2n})
\Gamma_{2n}(q,-q,p_3..,p_{2n})\delta(p_3+..+p_{2n})
\ee
and
\be
\int dx\left({\delta W\over\delta\p}\right)^2=\sum_{n,m}\Gamma_{2n}\circ
\Gamma_{2m}
\ee
where
$\Gamma_{2n}\circ
\Gamma_{2m}$ is

\bq
&& 8nm\pi 
\int dp_2..dp_{2n}dk_2..dk_{2m}\,
\tilde\p(p_2)..\tilde\p(p_{2n})\tilde\p(k_2)..\tilde\p(k_{2m})
\Gamma_{2n}(-(p_2+..+p_{2n}),p_2,..,p_{2n})
\nonumber\\
&&\,\,\,\times\Gamma_{2m}(-(k_2+..+k_{2m}),k_2,..,k_{2m})
\,\delta(p_2+..+p_{2n}+k_2+..k_{2m})
\eq
Expanding the $\Gamma$ in powers of $\hbar$ as
$\Gamma_{2n}=\sum\hbar^m\Gamma_{2n}^{\hbar^m}$
and ignoring order-$\hbar$ terms in the \Sc equation gives
the tree-level result 
\be
\Gamma^1_2\circ\Gamma^1_2+2\Gamma^1_2\circ\Gamma_4^1+2\Gamma_2^1\circ
\Gamma_6^1+\Gamma_4^1\circ\Gamma_4^1+..=\int dx\left(\p'^2+M^2\p^2+
{g\over 12}\p^4\right)
\ee
The term quadratic in $\p$ is
\be
\Gamma^1_2\circ\Gamma^1_2=\int dp 8\pi (\Gamma_2^1(p,-p))^2\tilde\p(p)
\tilde\p(-p)=\int {dp\over 2\pi}(p^2+M^2)\tilde\p(p)
\tilde\p(-p)
\ee
so if we take the negative root for normalisability of the vacuum
functional we get 
\be
\Gamma^1_2=-{\sqrt{p^2+M^2}\over 4\pi}\equiv -{\omega (p)\over 4\pi}.
\label{eq:twog}
\ee
Using
\be
\Gamma^1_2\circ\Gamma_{2n}=-\int dp_1..dp_{2n}
\tilde\p(p_1)..\tilde\p(p_{2n})\left(\sum_1^{2n}\omega(p_i)\right)
\Gamma_{2n}(p_1,..,p_{2n})\delta(p_1+..+p_{2n})
\ee
in  
\be
\Gamma^1_2\circ\Gamma^1_4
={g\over 4!}\int dx\p^4.
\ee
gives, for $p_1+..+p_4=0$, \cite{Hat},
\be
\Gamma^1_4(p_1,..,p_4)=-{g\over (2\pi)^3(4!)(\omega(p_1)+..+\omega(p_4))}.
\label{eq:gam4}
\ee
The terms of higher order in $\p$, for which there are no contributions
from the potential give $\sum_{n+m={\rm const}}\Gamma^1_{2n}\circ\Gamma^1_{2m}
=0$ which can be solved recursively as
\bq
&&\Gamma_{2r}^1(p_1,..p_{2r})=
\nonumber\\
&&\quad{4\pi\over\sum_1^{2r}\omega(p_i)}
\sum_{n=2}^{r-1} n(r+1-n){\bf S}\Big\{
\Gamma_{2n}^1(-(p_2+..+p_{2n}),p_2,..,p_{2n})\nonumber\\
&&\quad\quad\times\Gamma_{2(r+1-n)}^1(-(p_{2n+2}+..+p_{2(r+1-n)}),p_{2n+2},..,p_{2(r+1-n)})
\Big\}\label{eq:genr}
\eq
where ${\bf S}$ symmetrises the momenta.
Expanding $\Gamma^1_2$ and $\Gamma^1_4$ in 
positive powers of the momenta reproduces (\ref{eq:tree})
as it should since no re-summation is involved in 
either approach to the tree-level result.

\medskip
The order-$\hbar$ contribution to the \Sc equation is
\be
\sum_{n,m}\Gamma^1_{2n}\circ \Gamma^\hbar_{2m}+\sum_n\Delta
\Gamma_{2n}^1+\int dx \left(2{\cal E}^\hbar-(M^2)^\hbar \p^2\right)=0.
\ee
The term quadratic in $\p$ gives the limit as $\epsilon\rightarrow 0$ of
\be
{\delta M^2+M^2(\epsilon)^\hbar\over 4\pi}-2\omega(p)\Gamma_2^\hbar(p,-p)
+{g\over 32\pi^2}\int_{-1/\sqrt \epsilon}^{1/\sqrt \epsilon}
{dq\over \omega(q)+\omega(p)}=0
\label{eq:nn}
\ee
which can be solved for $p\neq 0$ as, \cite{Hat},
\be
\Gamma^\hbar_{2}(p,-p)={g\over 32\pi^2}\int_0^{\infty}
dq\left({1\over \omega(q)(\omega(q)+\omega(p))}\right)-
{\delta M^2\over 8\pi\omega (p)}
\nonumber
\ee

\be
={g\over 32\pi^2p}\sinh^{-1}\left({p\over M}\right)-
{\delta M^2\over 8\pi\omega (p)}
\label{eq:exp}
\ee
and for $p=0$ we get $\Gamma^\hbar_{2}=g/(32\pi^2M)-\delta M^2/(8\pi M)$.
The renormalisation condition that fixes $B_2$ at its classical value
requires that $\Gamma^\hbar_{2}(0,0)=0$, which determines $\delta M^2=g/(4\pi)$.
Setting $p=0$ in (\ref{eq:nn}) and taking the limit $\epsilon\rightarrow 0$
is meant to 
yield the 
same as taking the limit $\lambda\rightarrow \infty$ of 
$\tilde I^\hbar_0(\lambda)$ in (\ref{eq:io}) when we identify
$\Gamma_2(0,0)=B_2/(2\pi)$. This gives the value $-g/(8\pi)$
quoted earlier that agrees well with the large $\lambda$
behaviour of $S(\lambda)-\tilde M^2(\lambda)^\hbar/2$.
More particularly $S(\lambda)$ should be obtained from
the large $\epsilon$ expansion of
\be
H(\epsilon)={1\over 16\pi}\int_{q^2<1/ \epsilon}
{dq\over \omega(q)+\omega(p)}
\ee
by applying the two contour integral re-summations, giving
\be
-{\lambda\over 4\pi}\int_{|\tilde s|=\infty}d\tilde s\,\tilde s^{-3/4}
e^{\lambda^2\tilde s}\int_{|s|=\infty}ds\,s^{-1/2}e^{s/\sqrt{\tilde s}}
H(s)
\ee
which does in fact coincide with (\ref{eq:series}).
Since the large $\lambda$ behaviour corresponds to small
$\epsilon$ we can use (\ref{eq:exp}) to investigate this.
Thus for small $\epsilon$
\be
\int_0^{1/\sqrt\epsilon} dq\left({1\over \omega(q)(\omega(q)+M)}\right)=
\sqrt{M^{-1}+\epsilon}-
\sqrt\epsilon\approx {\sqrt M}-\sqrt\epsilon +{\epsilon\over 2\sqrt M}
-{\epsilon^2\over 8{\sqrt M}^3}+..
\ee
which leads to the power law corrections to the large-$\lambda$
behaviour described earlier. Expanding (\ref{eq:exp}) in positive powers of $p^2$
leads to the exact results for the $W_2^\hbar$ quoted earlier

\bq
&&
\int{dp\over 2\pi}\Gamma^\hbar_2(p,-p)\tilde\p(p)\tilde\p(-p)=
\nonumber\\
&&
-{g\over \pi}\int dx\Bigg({\p'^2\over 48}+{3\p''^2\over 160}
 +{ 15\p'''^2 \over 896}+{ 35\p''''^2\over 2304}+{ 315\p^{(5)2}\over
22528 }
\nonumber\\
&&\,\,\,\,\,\,\,
+{ 693\p^{(6)2}\over53248}
 +{ 1001\p^{(7)2}\over 81920 }+{6435 \p^{(8)2}\over 557056 }+{ 109395\p^{(9)2}
\over 9961472 }
 +{230945 \p^{(10)2}\over 22020096}+..\Bigg)
\eq

\bigskip
The $O(\hbar)$ contribution to the part of $W[\p]$ that is quartic 
in $\p$ is obtained from
\be
2\Gamma^1_2\circ\Gamma^\hbar_4+2\Gamma_4^1\circ\Gamma_2^\hbar+\Delta\Gamma_6^1=0
\ee
which leads to
\bq
&&
\Gamma^\hbar_4(p_1,..,p_4)=
{g^2\over (2\pi)^3 4!\pi\sum_1^4\omega(p_i)}{\bf S}\Biggl\{\int_0^\infty {dq
\over 2\omega(q)+\sum_1^4\omega(p_i)}\Biggl(-{1\over \omega(q)(\omega(q)+\omega(p_1))}
\nonumber\\
&&
+{3\over(\omega(q)+\omega(p_1)+\omega(p_2)+\omega(q+p_1+p_2))
(\omega(q)+\omega(p_3)+\omega(p_4)
+\omega(-q+p_3+p_4))}\Biggr)\nonumber\\
&&\hskip2in+{1\over 2\omega(p_1)\sum_1^4\omega(p_i)}\Biggl\}\label{eq:gam4h}
\eq
Expanding in positive powers of the momenta $p_1,..,p_4$ and 
integrating over $q$ numerically,
(using MAPLE), leads to 
\bq
&&W^\hbar_4={g^2\over 10000}\int dx \Biggl(3.973\p^4-19.45\p^2\p'^2
-7.961\p'^4         \nonumber\\
&&\,\,\,\,\,16.27\p^2\p''^2   +85.78\p'^2\p''^2 +33.66\p\p''^3-14.15\p^2\p'''^2
+..
\Biggr).
\label{eq:res4}
\eq
From this it is clear that our previous estimate was quite good, but that
the observation that the ratios $\rho_{j_0,..,j_n}=B^\hbar_{j_0,..,j_n}/(g
B^0_{j_0,..,j_n})$ are the same for coefficents of functionals of the same
dimension and number of fields is only approximate,
since 
\be
\rho_4=-0.03814
\ee
\be
\rho_{2,2}=-0.1245
\ee
\be
\rho_{0,4}=-0.2038,\rho_{2,0,2}=-0.2082
\ee
\be
\rho_{0,2,2}=-0.2834,\rho_{1,0,3}=-0.2872,\rho_{2,0,0,2}=-0.2898.
\ee
These ratios may be explained by observing that the dominant contribution
to (\ref{eq:gam4h}) comes from the last term
in the braces, which itself originates in the mass renormalisation. 
$\hbar\delta M^2$ was fixed by imposing $B^\hbar_2=0$, but if instead we had
taken $\hbar\delta M^2=0$ then this term would have been absent. The effect 
of this choice on the ratios can be calculated using dimensional analysis and gives
\be
\rho_4=-0.00165
\ee
\be
\rho_{2,2}=-0.00513
\ee
\be
\rho_{0,4}=-0.00486,\quad\rho_{2,0,2}=-0.00926
\ee
\be
\rho_{0,2,2}=-0.00488,\quad\rho_{1,0,3}=-0.00868,\quad\rho_{2,0,0,2}=-0.0113.
\ee
The advantage of this choice is that the one-loop corrections to
the coefficents $B_{j_0,..,j_n}$ for functionals containing four fields
are now significantly smaller.  The same is true for the coefficents
corresponding to two fields, with the exception of $B_2$. This suggests
that a more effective choice of renormalisation condition 
which would reduce the size of the one-loop corrections,
would be to
fix $B_4$ at its classical value, rather than $B_2$.
\vfill
\eject
\section{Sinh-Gordon Model}

From the standpoint of perturbation theory $\p^4$-theory is `close' to a theory that is
quite special, namely the Sinh-Gordon theory which has an infinite number of
conserved quantities that imply the absence of particle production, it is
interesting to calculate the vacuum functional for this case. 
The potential
of the Sinh-Gordon theory may be taken to be \cite{col}
\bq
&&V={M^2+\hbar\delta M^2\over\beta^2}\cosh (\beta\p)\,\exp\left(-{\beta^2\over 4\pi}
\int_0^{1/\sqrt\epsilon}{dp\over\omega(p)}\right)= \nonumber\\
&&(M^2+\hbar\delta M^2)\left({1\over\beta^2}+{\p^2\over 2}+{\beta^2\p^4\over 4!}+
{\beta^4\p^6\over 6!}+..\right)
\left(1-{\beta^2\over 4\pi}
\int_0^{1/\sqrt\epsilon}{dp\over\omega(p)}+..\right)
\eq
Apart from the replacement $g\rightarrow M^2\beta^2$ the Sinh-Gordon 
potential leads to the same expressions for the tree-level
values of $\Gamma^1_2,\Gamma^1_4$ and the one-loop result $\Gamma^\hbar_2$.
The tree-level $\Gamma^1_6$ is modified by the $\p^6$ term in the potential
\be
\Gamma_6^1\rightarrow\Gamma_6^1-{\beta^4M^2\over 6!(2\pi)^5 \sum_1^6\omega(p_i)}
\ee
this, together with the $\delta M^2\p^4$ term in $V$ modifies the one-loop
value $\Gamma^\hbar_4$
\be
\Gamma^\hbar_4\rightarrow \Gamma^\hbar_4 -
{\beta^4 M^2\over (2\pi)^4 4!\sum_1^4\omega(p_i)}\left(\int_0^\infty dq
\left({1\over 2\omega(q)+\sum_1^4\omega(p_i)}-{1\over 2\omega(q)}\right)
+{1\over 2}\right)
\ee
so that for the sinh-Gordon model
\bq
&&W^\hbar_4={\beta^4\over \pi}\int dx \Biggl({\p^4\over 384}+{5\pi-22\over1280}\p^2\p'^2
-{2275\pi-8952\over860160}\p'^4     
+{651\pi-2768\over 172032}\p^2\p''^2     \nonumber\\
&&\,\,\,\,\,  -{1041705\pi-4243072\over 41287680}\p'^2\p''^2 -{689535\pi-2920448\over 82575360}\p\p''^3\nonumber\\
&&\,\,\,\,\,
+{13905\pi-58624\over 3932160}\p^2\p'''^2
+..
\Biggr)\nonumber\\
&&\simeq{\beta^4\over 10000}\int dx \Biggl(8.2893\p^4-15.647\p^2\p'^2
-6.6791\p'^4         \nonumber\\
&&\,\,\,\,\,+13.3743\p^2\p''^2   +74.818\p'^2\p''^2 +29.0731\p\p''^3-12.0941\p^2\p'''^2
+..
\Biggr).
\eq
The ratios of the one-loop coefficents to their tree-level values
for this model are
\be
\rho_{4}=-0.07958
\ee
\be
\rho_{2,2}=-0.1001
\ee
\be
\rho_{0,4}=-0.1710,\quad\rho_{2,0,2}=-0.1712
\ee
\be
\rho_{0,2,2}=-0.2471,\quad\rho_{1,0,3}=-0.2481,\quad\rho_{2,0,0,2}=-0.2477.
\ee
Note that again the ratios are approximately the same for 
coefficents of functionals of the same number of fields and dimension,
however this cannot be explained away as simply the effect of
mass or coupling renormalisation. If we had taken $\delta M^2=0$
we would have obtained
\be
\rho_{4}=-0.1194
\ee
\be
\rho_{2,2}=-0.06035
\ee
\be
\rho_{0,4}=-0.05162,\quad\rho_{2,0,2}=-0.05183
\ee
\be
\rho_{0,2,2}=-0.04820,\quad\rho_{1,0,3}=-0.04915,\quad\rho_{2,0,2}=-0.04875.
\ee
which are approximately constant for coefficents of functionals of the same number of fields and dimension. These coefficents would, with
a change of sign, apply to the Sine-Gordon model as well.
\section{Feynman Diagram Expansion}
\begin{figure}[h]
  \centerline{
    \epsfxsize=6.0in
    \epsffile{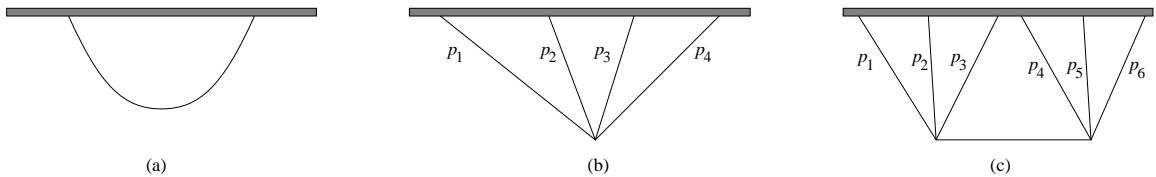}
             }
  \caption[tree]{\label{tree}Tree Diagrams up to $\varphi^6$ }
\end{figure}

We will now describe how the results of the previous 
section are obtained within the more conventional 
approach to field theory based on Feynman diagrams.
There are several ways to represent the vacuum functional as a functional integral, 
the most convenient for our purposes is the representation,
\cite{Sym},
\be
\Psi [\varphi ] = \int {\cal D}\phi e^{-S_E[\phi] + \int dx \ \varphi {\dot \phi}}
\label{eq:fint} 
\ee
where $\phi$ vanishes on the surface $t = 0$, which is the boundary of the 
Euclidean space-time $t < 0$ in which the
field $\phi$ lives. Therefore, $W[\varphi]$ is a sum of connected Euclidean Feynman 
Diagrams in which $\varphi$ is the source
for ${\dot \phi}$ on this boundary, the only major difference from the usual Feynman 
diagrams encountered in free space is that the propagator vanishes when either of 
its arguments lies on the boundary. Such a propagator can be obtained using the method
of images as
\be
G(\xx ,\yy) = G_0(\xx ,\yy) -  G_0(\xx ,{\underline \yy}) = G_0(\xx ,\yy) -  
G_0({\underline \xx } ,\yy)
\ee
where $\xx = (t,x)$, ${\underline \xx } = (-t,x)$, similarly for $\yy$
and $\underline\yy$, and $G_0(\xx ,\yy)$ is the free space propagator. 
Developing the Feynman diagram expansion yields the tree level diagrams up to $\varphi^6$ which are shown in figure 
(\ref{tree}).

The Feynman rules in the coordinate space for the diagram $(b)$, for instance, yield
\be
g s_f \int dx_1..dx_4  d^2{\bf v} \, \varphi(x_1)..  \varphi(x_4)   \ \frac{\del G(\xx_1, {\bf v})}
{\del t_1} \frac{\del G(\xx_2, {\bf v})}{\del t_2} \frac{\del G(\xx_3, 
{\bf v})}{\del t_3} \frac{\del G({\xx_4}, {\bf v})}{\del t_4} \label{eq:consp}
\ee
where $s_f$ is the  symmetry factor associated to the diagram, the times $t_1,..,t_4$ 
are all set to zero, and the time-like component of $\bf v$ is
integrated over negative values only. When a propagator ends on the boundary 
it is equal to twice the free space propagator, so that in momentum space
this is proportional to
\be
\int\,dp_1..dp_4\,dq_1..dq_4\tilde\p(p_1)..\tilde\p(p_4)
\delta(p_1+..+p_4)\,\delta(q_1+..+q_4)\prod_{j=1..4}{q_j\over p^2_j+q^2_j+M^2}
\ee
The momenta $q_1,..,q_4$ are the time-like components which, when
integrated over, give rise to $\delta$-functions that
impose the vanishing of $t_1,..,t_4$ in the configuration space
representation (\ref{eq:consp}). In the momentum-space representation these 
integrals are 
readily performed as contour integrals leading to our
previous expression (\ref{eq:gam4}) for $\Gamma_4^1$.
Diagram $(a)$ gives
\be
\frac{2}{(2 \pi)^2} \int dp dq \ {\tilde \varphi}(p) {\tilde \varphi}(- p) 
\frac{q^2}{p^2 + q^2+M^2}
\label{gamma2} 
\ee
The integral over $q$ is divergent. Formally we can write it as
\be
  \int \frac{dp}{2 \pi} \ {\tilde \varphi}(p) {\tilde \varphi}(-p) 
\left(\delta (0)-\sqrt{p_2^2 + M^2}\right) 
\ee
The origin of $\delta (0)$ is in the construction of the original path integral representation of $W$, 
\cite{Paul1}. Rather than the functional integral (\ref{eq:fint}),
we should start with 
\be
\Psi [\varphi ] = \langle \varphi | 0 \rangle =  \langle D | e^{i \int {\hat \pi} \varphi \ dx} | 
0 \rangle \ 
\label{eq:op}
\ee
(where the state $\langle D |$ satisfies $\langle D | \ {\hat \varphi} = 0$) . As
operators ${\hat \pi} = \dot {\hat \phi}$ , but in the passage from (\ref{eq:op}) to the functional
integral
${\hat \pi}$ is represented by $\dot \phi$ plus terms coming from the time derivative 
acting on the $T$-ordering because the functional integral represents $T$-ordered products. 
This leads to terms like $\int dx\p^2\delta(0)$,
because this is local it may be cancelled by an equal but opposite counterterm,
which amounts to simply discarding the divergence. Alternatively, and perhaps more satisfactorily, 
we can deal with this divergence by placing the 
source terms $\hat \pi \varphi$ not at $t = 0$ but at small, distinct times $t_i$ and 
finally taking the limit $t_i \rightarrow 0$ .
This leads to an extra factor of the form $e^{i q \epsilon}$ in (\ref{gamma2}) 
which regularizes the divergence and enables the integral to be done
by closing the $q$-contour in the upper half-plane, yielding a finite
result in limit as $\epsilon \rightarrow 0$, so that we
end up with our earlier expression (\ref{eq:twog}).

  When neither end of a propagator is on the boundary, $t=0$, the image
charge breaks energy conservation leading to more
complicated expressions. However for the tree-level diagram $(c)$ 
these may be combined into an expression
proportional to
\be
\int \ \left( \prod_1^6 \frac{dp_{i} dq_{i} \ {\tilde \varphi}(q_i) \ p_i
}{p_i^2 + q_i^2 + M^2}\right) \
\frac{\delta (p_1 +..+ p_6)  \  \delta (q_1 +..+ q_6)} {(p_1 + p_2 +
p_3)^2 + (q_1 + q_2 + q_3)^2 + M^2}
\ee
which, after integration over the $q_i$ gives the same as (\ref{eq:genr}).

\begin{figure}[h]
  \centerline{
    \epsfxsize=6.0in   
    \epsffile{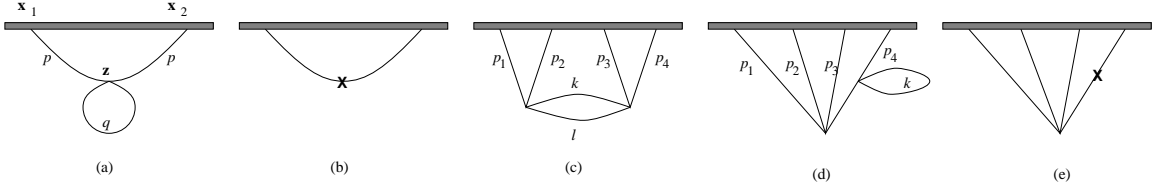}
             }
  \caption[bubble]{\label{bubble} Two and four field one loop diagrams}
\end{figure}

The diagrams represented in figure (\ref{bubble}) enable us to calculate the $O(\hbar)$ 
correction for the coefficients of
$2$ and $4$ fields .

From fig.\ref{bubble}(a) above we can work out the first order correction to $B_2$ as \be
s_f \ g \int dx_1 dx_2 d^2{\bf z} \ \varphi (x_1) \varphi (x_2) \ \frac{\partial G 
({\bf x}, {\bf z})}{\partial t_1} G( {\bf z}, {\bf z})  \frac{\partial G ({\bf y}, {\bf z})}{\partial 
t_2} 
\label{massren}
\ee 
This is divergent due to the factor $G( {\bf z}, {\bf z})$.
Since the propagator does not touch the boundary, it has an
energy non-conserving 
contribution from the image represented by the exponential in
\be
G ( {\bf z}, {\bf z}) = \int \frac{dp\,dq}{(2 \pi)^2} \ \frac{(1 - e^{2 i q t})}{(p^2+q^2 + M^2)} 
 \ . 
\ee
where $t$ is the time-like coordinate of z. We regulate the integral by restricting the space-like 
component of
momentum, $p^2<1/\epsilon$, just as we did for the Laplacian
in the Hamiltonian. The divergence is then cancelled by
the counter-term represented by fig.\ref{bubble}(b),
which is due to the order-$\hbar$ terms in $M^2(\epsilon)$, (\ref{eq:M}).
Taken together these diagrams give $\Gamma^\hbar_2$ as in (\ref{eq:exp}).

The first order correction $\Gamma_4^\hbar$ is given by the sum of
diagrams $\ref{bubble}(c)$,  $\ref{bubble}(d)$ and  $\ref{bubble}(e)$. As before, diagram $(e)$ is the 
counterterm diagram associated to the bubble appearing in  $(d)$. In
momentum space $\ref{bubble}(c)$ gives
\bq
&&{g^2 s_f^{(c)} \over 8 \pi^4} \ \int dk \,dp_1..dp_4
 \ {\tilde \varphi}(p_1)..{\tilde \varphi}(p_4)
 \  \delta (p_1+..+p_4) \nonumber \\
&&   \frac{\sum_1^4\omega(p_i) + \omega(k) + \omega(l)}{\sum_1^4\omega(p_i) (\omega (p_3) + \omega(p_4) + \omega(k) 
+\omega(l))(\omega(p_1) + \omega(p_2) +  \omega(k) + \omega(l) )} . \nonumber \\
&& \frac{1}{(\sum_1^4\omega(p_i) + 2 \omega(k) )(\sum_1^4\omega(p_i) + 2 \omega(l) )} ,
\eq 
where $l=p_3+p_4+k$.
whereas, for the sum of the other two diagrams, (with $\delta M^2=0$), we have
\be
 - {4g^2 s_f^{(d)} \over{ \pi^4}} \ \int dk \,dp_1..dp_4
 \ {{\tilde \varphi}(p_1)..{\tilde \varphi}(p_4)
\  \delta (p_1 +.. + p_4) \over
 {\omega(k) (\omega(p_4) + \omega(k))\sum_1^4\omega(p_i) (\sum_1^4\omega(p_i) + 2 \omega(k))}}
\ee
The symmetry factors $s_f^{(c)}$ and $s_f^{(d)}$ are respectively equal to  $1/16$ and $1/12$. 
If we expand in powers of the momenta $p_1,..,p_4$ 
the loop integration over $k$ can be done to reproduce the results for
$W^\hbar_4$ given by (\ref{eq:res4}).

\begin{figure}[h]
  \centerline{
    \epsfysize=1.2in
    \epsffile{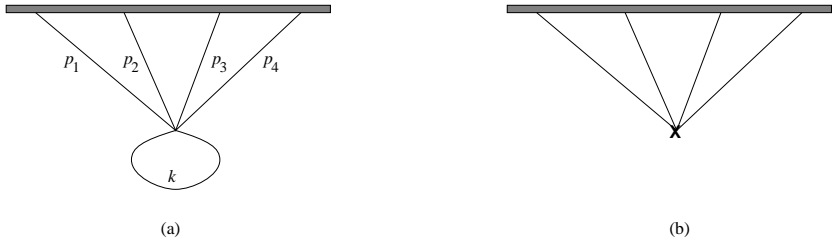}
             }
  \caption[sinegord]{\label{sinegord} $6$ point interaction for the Sinh-Gordon Model}
\end{figure}

The extra $\p^6$ term in the potential of the Sinh-Gordon model 
generates the diagram fig. (\ref{sinegord}).
Its  analytic expression after subtracting the counterterm reads
\be
\beta^4 \ s_f \frac{1}{(2 \pi)^3} \int dk \,dp_1..dp_4
 \ {{\tilde \varphi}(p_1).. {\tilde \varphi}(p_4) 
 \delta (p_1 + ..+p_4)\over
 {\omega (k) (2 \omega(k) + \sum_1^4\omega(p_i))} }\ .
\ee
where $s_f$ for this diagram is $1/96$. This gives the 
modifications to the $\p^4$ results described earlier.

\section{Reconstructing the Vacuum Functional}

\begin{figure}[h]
\unitlength1cm  
\begin{picture}(14,10)
\put(8.5,8.4){$\sqrt{p^2+1}$}
\put(9,6.4){$S_{12}$}
\put(4.6,7){$C$}
\epsfysize=40cm
\epsffile{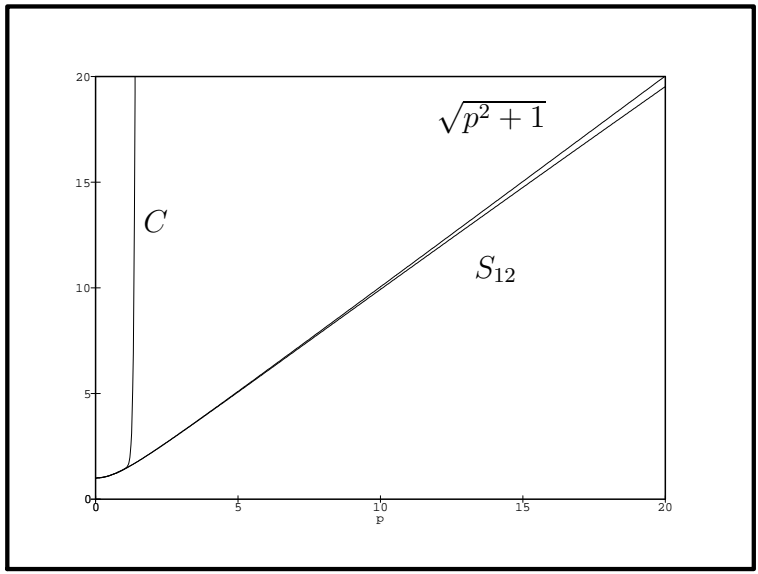} 
\end{picture}
\caption[wave2]{\label{wave2} Tree-level vacuum functional:
$\Gamma^1_2(p,-p)$}
\end{figure}

We will now use the above results for the $\p^2$
part of $W[\p]$ to illustrate how the vacuum functional
can be reconstructed from its large distance expansion.
The tree-level contribution has been treated in detail in 
\cite{Paul1}. Applying formula (\ref{eq:resss}) to
$\int dp\tilde\p (p)\tilde\p (-p)\Gamma_2(p,-p)$
amounts to the expression
\be
\Gamma_2(p,-p)=\lim_{\lambda\rightarrow\infty}{1\over 2\pi i}\int_{|s|=\infty}
{ds\over s-1}e^{\lambda(s-1)}\sqrt{s}\Gamma_2(p/\sqrt{s},-p/\sqrt{s})
\ee
Since $|s|$ is large on the contour we can use the local
expansion $\Gamma_2(p,-p)=\sum_0^\infty  a_n p^{2n}$. Shifting
$s$ we get
\be
\lim_{\lambda\rightarrow\infty}{1\over 2\pi i}\int_{|s|=\infty}
{ds\over s}e^{\lambda s}\sqrt{s+1}\sum_0^\infty  a_n {p^{2n}
\over (s+1)^n}\equiv \lim_{\lambda\rightarrow\infty} S(p,\lambda)
\ee
Expanding the $(s+1)$ factors in powers of $1/s$ enables
the integral to be done, yielding a power series in $\lambda$.
For example, at tree-level
\be
{1\over 2\pi i}\int_{|s|=\infty}
{ds\over s-1}e^{\lambda(s-1)}\sqrt{p^2+s}
=
\sum_0^\infty {(-)^{n+1}\lambda^{n-1/2}(1+p^2)^n
\over n!(2n-1)\sqrt \pi}.\label{eq:wtre}
\ee
This series converges for all positive $\lambda$.
We get an approximation by truncating the expansion 
by including terms up to and including $\lambda^{N-1/2}$,
say. This requires a knowledge of the local expansion only
up to terms in $(\p^{(N)}(x))^2$. To demonstrate this
approximation we have plotted in fig. (\ref{wave2}) the
series (\ref{eq:wtre}) truncated at $N=12$, $S_{12}$. 
The value of $\lambda$
is chosen so that the last term included is one per cent
of the value of the series. We have also plotted $\sqrt{p^2+1}$,
and the expansion of $\sqrt{p^2+1}$ in powers of $p^2$, $C$
truncated to fourteen terms. The full series
fails to converge for $p^2>1$, and this is
reflected in the fact that the truncated series ceases
to be a good approximation for $p^2>1$. However,
$S_{12}$, which is a resummation of this series is
a very good approximation for a much larger range 
of momenta. 

\begin{figure}[h]
\unitlength1cm  
\begin{picture}(14,10)
\put(8.5,7.8){$ S_{12}$}
\put(10.5,6.4){$R_{12}$}
\put(9,5.4){arcsinh(p)/p}
\put(5,7){$C$}
\epsfysize=40cm
\epsffile{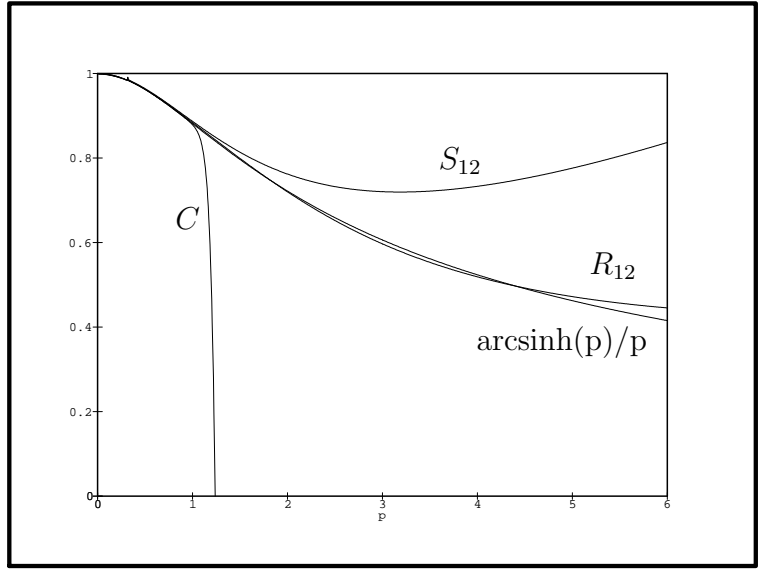} 
\end{picture}
\caption[wave1]{\label{wave1} One-loop vacuum functional: $\Gamma^\hbar_2(p,-p)$}
\end{figure}

The one-loop correction, $\Gamma_2^\hbar$, may be treated in the same way.
In fig. (\ref{wave1}) we have plotted ${\rm arcsinh}(p)/p$.
The small $p$ expansion, $C$ is again only good for $p^2<1$.
Our approximation that re-sums this series, $S_{12}$ provides
a good approximation only over a slightly larger range.
The accuracy of this approximation is greatly improved
by further re-summations, just as we did for the Laplacian.
Let us define the re-summation operator acting on a function
of $\lambda$ and $p$ to be
\be
R\cdot S(\lambda,p)\equiv
{1\over 2\pi i}\int_{|s|=\infty}
{ds\over s}e^{\lambda s} S(s^{-1/2},p).
\ee
Then the curve $R_{12}$ shown in fig. (\ref{wave1})
results from applying $R$ twice to $S_{12}$,
and provides a good approximation to
${\rm arcsinh}(p)/p$ for values of $p$ up to 
about $p=5$. Since the effect of applying $R^p$ to a term
in $S_{12}$ that is proportional to $\lambda^n$
is simply to divide it by
$\Gamma(n/2+1)\Gamma(n/4+1)..\Gamma(n/2^p+1)$ further 
applications would have no significant effect 
when we take just twelve terms in the expansion.

\section{Conclusions}
The purpose of this paper has been to test an approach to
quantum field theory in which states are constructed 
in the \Sc representation from their
large-distance behaviour. The vacuum functional 
is expanded as a series
of local functionals. Truncating this series at some convenient
order leads to an approximation scheme. For 
scalar $\p^4$ theory in 1+1-dimensions
we compared the results obtained
by solving the form of the \Sc equation that applies 
in this approach with those obtained from a semi-classical
expansion, and from the Feynman diagram expansion of the wave-functional.
We have found that the expansion coeficents agree to within a few per cent 
when we truncate these series at about ten terms. 
Also we found that the known counterterms that contain information
about short-distance effects are correctly reproduced
by our large-distance expansion to a similar accuracy.

We also found a curious simplification that occurs in the vacuum
functional of the Sinh-Gordon and Sine-Gordon models.
The ratios of coefficents of the one-loop corrections to the coefficents of
local functionals containing four fields to their tree-level values
are approximately the same for functionals of the same dimension. 

\section{Acknowledgements}

M. Sampaio acknowledges a grant from  CNPq - Conselho Nacional de Desenvolvimento 
Cient\'{\i}fico e Tecnol\'{o}gico - Brasil, and J. Pachos 
acknowledges a studentship from the University of Durham.

\end{document}